\documentclass[11pt]{article}
\usepackage{amsmath}
\usepackage{amsfonts}
\usepackage{amssymb}
\usepackage{epsfig}
\usepackage{times}
\newcommand{\be}{\begin{equation}}
\newcommand{\ee}{\end{equation}}
\newcommand{\ba}{\begin{array}}
\newcommand{\ea}{\end{array}}
\newcommand{\bea}{\begin{eqnarray}}
\newcommand{\eea}{\end{eqnarray}}

\def\ket#1{\,\left\vert  #1\right\rangle}
\def\bra#1{\left\langle#1\right\vert\, }
\def\vac{\ket{0}} 
\def\b{\rm b}
\def\a{\rm a}
\def\D{\rm D}
\def\c{\rm c}
\def\d{\rm d}
\def\dpsi{[\D \psi]}
\def\Dpsi{[\D \overline{\psi}]}
\def\dchi{[\D \chi]}
\def\Dchi{[\D \overline{\chi}]}
\def\dPhi{[\D \Phi]}
\def\DPhi{[\D \overline{\Phi}]}

\def\E{\rm exp}

\def\hbar{\not{\hbox{\kern-2.3pt $h$}}}
\def\psl{\not{\hbox{\kern-2.3pt $p$}}}
\def\Psl{\not{\hbox{\kern-2.3pt $P$}}}
\def\ksl{\not{\hbox{\kern-2.3pt $k$}}}
\def\qsl{\not{\hbox{\kern-2.3pt $q$}}}
\def\slad{\not{\hbox{\kern-2.3pt $\partial$}}}
\def\I{\rm i}
\begin{document}
\begin{titlepage}

October 1999 (revised November 1999) \hfill PAR-LPTHE 99/35
\vskip 4.5cm
{\baselineskip 17pt
\begin{center}
{\bf  ON THE DYNAMICAL SYMMETRY BREAKING OF THE ELECTROWEAK INTERACTIONS BY THE TOP QUARK}
\end{center}
}

\vskip .5cm
\centerline{
Nguyen Van Hieu
     \footnote[1]{E-mail: nvhieu@ims.ncst.ac.vn}
}
\vskip 2mm
\centerline{
\em Institut of Physics, NCST. Hanoi
     \footnote[2]{PO Box 429, Bo Ho, Hanoi 10000, (Vietnam)}}
\vskip 3mm
\centerline{
Pham Xuan Yem
     \footnote[3]{E-mail: pham@lpthe.jussieu.fr}
            }
\vskip 2mm
\centerline{{
\em Laboratoire de Physique Th\'eorique et Hautes Energies, Paris}
     \footnote[4]{LPTHE tour 16\,/\,1$^{er}\!$ \'etage,
          Universit\'e P. et M. Curie, BP 126, 4 place Jussieu,
          F-75252 PARIS CEDEX 05 (France).}
}
\centerline{\em Universit\'es Paris 6 et Paris 7;} \centerline{\em Unit\'e associ\'ee au CNRS, UMR 7589.}
\vskip 1.5cm
{\bf Abstract:} We discuss the electroweak gauge symmetry breaking 
triggered by a new strong attractive  interaction to condensate fermion-antifermion, and topcolor is a prototype.  To deal with the fermion 
pairing, a general method based on the Hubbard-Stratonovich transformation in the functional integral approach is used. 

We derive a formula which relates  the $W^\pm$, $Z^0$  weak boson masses 
to that of the condensated fermion, thus generalizing the Pagels-Stokar formula obtained in QCD. The custodial $SU(2)$ electroweak symmetry turns out to be systematically violated, the deviation of $\rho\equiv M_W^2/(M_Z^2\cos^2\theta_W) $ from unity is related to the new physics scale $\Lambda$. Some phenomenological consequences of the top-pair condensation models are discussed. Distinctive signatures of the $\overline{t}t$ scalar bound state, a Higgs boson like denoted by $H_t$, are the dominant decay modes  $H_t\to \Upsilon +\gamma\;,$ $H_t\to \Upsilon +Z^0$,  and $H_t \to B^* +\overline{B}^*$.

\smallskip

{\bf PACS} number(s): 11.15.Ex \quad 11.15.Tk \quad 12.60.Fr \quad 14.80.Cp
\vfill
\null\hfil\epsffile{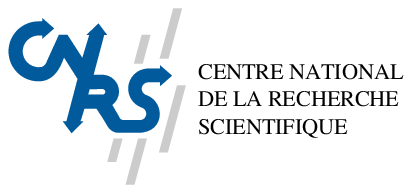}
\end{titlepage}
%
%
 The well-known Higgs mechanism\cite{Higgs, Eng, Gur} that involves an elementary scalar 
field may not  be the unique scenario to spontaneously break the gauge 
symmetry of the standard electroweak theory\cite{Gsw}. A dynamical 
symmetry breaking (DSB) due to the condensation of some fermion-antifermion 
pair may also generate masses to the gauge bosons, a typical example borrowed from superconductivity is the Cooper electron pair. Another example  is the nonzero vacuum expectation value of massless quark-antiquark pair, its condensate breaks the QCD chiral symmetry. In all cases,  nonzero numbers must be assumed to break the symmetries, i.e. $\bra 0 \Phi  \vac \neq 0$, 
$\bra 0 $ e$^-$ e$^- \vac \neq 0$,
$\bra 0 \overline{q} q \vac \neq 0$ respectively in the standard Higgs mechanism, the Cooper electron pair in superconductivity, and quark-antiquark pair 
in QCD.

With DSB, in order to have large values for the W$^{\pm}$ and Z$^0$ weak boson masses,  
there must exist beyond the standard model some new 
attractive interaction with sufficiently massive fermions involved which replaces the Higgs potential $ \lambda \Phi^4 +\mu^2 \Phi^2$ with the wrong sign $\mu^2<0$.  
This idea has  motivated 
the topcolor interaction\cite{Nambu, Mir, Bar, yem} and its extension\cite{Hil, Del, Chi} as the dynamical breaking of the  
electroweak  symmetry due to the condensation of the top-antitop pair since the top quark is the most massive elementary particles. A review  of the top-condensation models with extensive references is recently 
available\cite{Cvetic}.    

An attempt is made in this note to establish a relation between the 
masses of the condensated fermion and the weak vector bosons $W^\pm$, $Z^0$. The  Hubbard-Stratonovich transformation\cite{Hub} applied to the functional integral method, previously developped by one of us\cite{hieu} for condensed matter physics, turns out to be particularly powerful for treating the problem of fermion pairing  considered here.

We start by introduce a system of left-handed top and bottom quarks put into an SU(2) doublet:
\begin{equation}
\psi_{\a},\; \a=1,2 \;\; {\rm with}\;\;\psi_1(x) ={1-\gamma_5\over 2}\psi_b(x) \,,\;\;\;\; \psi_2(x) ={1-\gamma_5\over 2}\psi_t(x)\;,
\end{equation}
and a singlet fermion $\chi$ which is the right-handed top quark,
\begin{equation}\chi(x)= {1+\gamma_5\over 2}\psi_t(x)\;.
\end{equation}
We further postulate that the effective four-fermion topcolor interaction -- mediated by topgluons ${\cal G}_t$ -- may be written in the form
\begin{equation}
{\cal L}_{int} =\chi_{\delta}(x)\overline{\psi}^{\a,\alpha}(x) V_{\alpha \gamma}^{\delta \beta} \psi_{\a, \beta}(x) \overline{\chi}^{\gamma}(x)\;,
\end{equation}
\begin{equation}
 V_{\alpha \gamma}^{\delta \beta}  = G (\gamma_\mu)_{\alpha}^{\beta}
(\gamma^\mu)_{\gamma}^{\delta}
\end{equation}
with some strong coupling constant $\,G\,$ having the (mass)$^{-2}$ dimension. The quark color index is implicitly understood, however it is convenient to explicit the spinor indices $\alpha, \cdots, \delta =1\cdots 4$. Starting from massless fields, the role of the topcolor interaction (4) is to dynamically generate masses to both the top quark  as well as to the gauge bosons. On general ground, one would expect that these masses are functions of the coupling constant $G$ and of the  {\it new physics} energy scale $\Lambda$, as we will see later.
We consider now the functional integral of the system

$$ Z = \int \dpsi \; \Dpsi\; \dchi\; \Dchi \;\;\E \left\{\I \int \d^4 x 
\left[\overline{\psi}^{\a,\alpha}(x)
(\slad)_{\alpha}^{\beta} \psi_{\a, \beta}(x) + \overline{\chi}^{\alpha}(x) 
(\slad)_{\alpha}^{\beta} \chi_{\beta}(x)\right] \right\}$$
\begin{equation}
 \times \;\E \left\{\I \int \d^4 x \;\;\chi_{\delta}(x)
\overline{\psi}^{\a,\alpha}(x) V_{\alpha \gamma}^{\delta \beta}\psi_{\a, \beta}(x) \overline{\chi}^{\gamma}(x) \right\}\;.
\end{equation}
For the free fermion system without $V_{\alpha \gamma}^{\delta \beta}$ interaction, the functional integral becomes
\begin{equation}
Z_0 = \int \dpsi \; \Dpsi\; \dchi\; \Dchi \;\;\E \left\{\I \int \d^4 x 
\left[\overline{\psi}^{\a,\alpha}(x)
(\slad)_{\alpha}^{\beta} \psi_{\a, \beta}(x) + \overline{\chi}^{\alpha}(x) 
(\slad)_{\alpha}^{\beta} \chi_{\beta}(x)\right] \right\}\;.
\end{equation}
Let us introduce\cite{hieu} the dimensional (mass)$^3$ auxiliary fields denoted by $\Phi^{\gamma}_{\a, \alpha}(x)$ and $ \overline{\Phi}^{\a, \alpha}_{\gamma}(x)$ which represent the fermion-antifermion system, where $\overline{\Phi}$ is 
defined from $\Phi$ as  follows:
$$ \overline{\Phi}^{\a, \alpha}_{\gamma}(x) =(\gamma_0)^\delta_\gamma \;\Phi^{\dagger \a, \beta}_\delta (x) \; (\gamma_0)^\alpha_\beta =(\gamma_0)^\delta_\gamma 
\left(\Phi^{\delta}_{\a, \beta}(x)\right)^* (\gamma_0)^\alpha_\beta \;.$$ 
The associated functional integral for these auxiliary fields is 
\begin{equation}
Z_0^{\Phi} = \int \dPhi \; \DPhi\;  \;\;\E \left\{-\I \int \d^4 x 
\;\overline{\Phi}^{\a,\alpha}_{\delta}(x)V_{\alpha \gamma}^{\delta \beta}
\Phi^{\gamma}_{\a, \beta}(x) \right\} \;.
\end{equation}
 Now we apply the Hubbard-Stratonovich transformation\cite{Hub} to the interacting part of the action (3) in the functional integral, and get
$$
 \E \left\{\I \int \d^4 x \,\chi_{\delta}(x)
\overline{\psi}^{\a,\alpha}(x) V_{\alpha \gamma}^{\delta \beta}\psi_{\a, \beta}(x) \overline{\chi}^{\gamma}(x) \right\} = {1\over Z_0^{\Phi} } \int \dPhi \DPhi\; \E \left\{-\I \int\d^4 x \,\overline{\Phi}^{\a,\alpha}_{\delta}(x)V_{\alpha \gamma}^{\delta \beta}
\Phi^{\gamma}_{\a, \beta}(x) \right\}$$    
\begin{equation}
\times \; \E \left\{ -\I \int \d^4 x \left[\overline{\Delta}_\gamma^{\a, \beta}(x)
 \psi_{\a, \beta} (x) \overline{\chi}^\gamma (x) + \chi_\delta (x) 
\overline{\psi}^{\a, \alpha} (x) \Delta^\delta_{\a, \alpha} (x)\right] \right\} \;,
\end{equation}
where
\begin{equation}
\Delta^\delta_{\a, \alpha} (x) =  V_{\alpha \gamma}^{\delta \beta} \Phi^{\gamma}_{\a, \beta}(x) \;\;,\;\;
\overline{\Delta}^{\a,\beta}_{\gamma}(x) = \overline{\Phi}^{\a,\alpha}_{\delta}(x)V_{\alpha \gamma}^{\delta \beta} \;.
\end{equation} 
Physically, the $\Delta^\delta_{\a, \alpha}(x)$ defined above represents a bosonic field which is  a bound state of the fermion-antifermion pair due to the strong interaction $V_{\alpha \gamma}^{\delta \beta}$. It is not necessarily a scalar field and it has the canonical (mass)$^1$ dimension.

Substituting the expression (8) into the r.h.s. of (5) and integrating out over all the fermionic functional variables $\psi_{\a}, \overline{\psi}^{\a}, \chi,
\overline{\chi}\,$, we can express $Z$ as a functional integral over only the auxiliary fields
$ \Phi^{\gamma}_{\a, \beta}(x)$ and $\overline{\Phi}^{\a,\alpha}_{\gamma}(x)$, thus
\begin{equation}
Z= {Z_0 \over Z^{\Phi}_0} \int  \dPhi \; \DPhi\; \E \left( \I S_{eff} 
\left[\Phi,\overline{\Phi} \right] \right)
\end{equation}
with some effective action\cite{hieu}
\begin{equation}S_{eff} 
\left[\Phi,\overline{\Phi} \right]  = -\int \d^4 x \;\overline{\Phi}^{\a,\alpha}_{\delta}(x) \;V_{\alpha \gamma}^{\delta \beta} \Phi^{\gamma}_{\a, \beta}(x)  +\sum_{n=1}^{\infty} W^{(2n)} \left[ \Delta,\overline{\Delta}\right]\;,
\end{equation}
and $W^{(2n)} \left[ \Delta,\overline{\Delta}\right]$ is a functional of the $n$-th order with respect to each kind of fields 
$\Delta^\delta_{\a, \alpha} (x)$ and $\overline{\Delta}^{\a,\beta}_{\gamma}(x)$.
In order to write $W^{(2n)}\left[ \Delta,\overline{\Delta}\right]$ in a compact form, we introduce the $4\times 4$ matrices $\widehat{\Delta}_{\a} (x)$ and 
$\widehat{\overline{\Delta}^{\a}} (x)$ with the elements
$$\left[\widehat{\Delta}_{\a} (x)\right]^\gamma_\alpha = \Delta^\gamma_{\a, \alpha} (x) \;\;,\;\; \left[\widehat{\overline{\Delta}^{\a}} (x)\right]_\gamma^\alpha 
= \overline{\Delta}_\gamma^{\a, \alpha} (x) \;.$$%
Then we have
\begin{equation}
W^{(2)} \left[ \Delta,\overline{\Delta}\right] = \I \int \d^4 x \; \d^4 y \;Tr\left[
\widehat{\overline{\Delta}^{\a}} (x) S^L(x-y)\widehat{\Delta}_{\a} (y)
S^R(y-x)\right]\;,
\end{equation}

$$W^{(4)} \left[ \Delta,\overline{\Delta}\right] = {\I\over 2} \int \d^4 x_1 \; \d^4 y_1 \;\d^4 x_2 \; \d^4 y_2 \;Tr\left[
\widehat{\overline{\Delta}^{\a_1}} (x_1) S^L(x_1-y_1)\widehat{\Delta}_{\a_1} (y_1)
S^R(y_1-x_2)\right.$$
\begin{equation}
\times\;\left.\widehat{\overline{\Delta}^{\a_2}}(x_2) S^L(x_2-y_2)\widehat{\Delta}_{\a_2} (y_2)
S^R(y_2-x_1)\right] \;,
\end{equation}

$$W^{(2n)} \left[ \Delta,\overline{\Delta}\right] = {\I\over n} \int \d^4 x_1 \; \d^4 y_1 \cdots\d^4 x_n \; \d^4 y_n \;Tr\left[
\widehat{\overline{\Delta}^{\a_1}} (x_1) S^L(x_1-y_1)\widehat{\Delta}_{\a_1} (y_1)
S^R(y_1-x_2) \cdots \right.$$
\begin{equation}
\times\;\left. \cdots\widehat{\overline{\Delta}^{\a_n}} (x_n) S^L(x_n-y_n)\widehat{\Delta}_{\a_n} (y_n)
S^R(y_n-x_1)\right] \;,
\end{equation}

where $ S^L(x-y)$ and  $ S^R(x-y)$ being respectively the propagators of left-handed and right-handed massless fermions,
\begin{equation}
\slad S^L(x-y) ={1-\gamma_5\over 2} \delta(x-y)  \;\;,\;\;
\slad S^R(x-y) ={1+\gamma_5\over 2} \delta(x-y) \;.
\end{equation}
From the expression (11) of the  auxiliary fields effective action, we derive the field equations 
\begin{equation}
\Delta^\delta_{\a, \alpha} (x) = V_{\alpha \gamma}^{\delta \beta} \sum_{n=1}^{\infty} {\partial \;W^{(2n)} \left[ \Delta,\overline{\Delta}\right]\over \partial \; 
\overline{\Delta}^{\a, \beta}_\gamma (x)}\;.
\end{equation}
The bosonic fields $\Delta^\delta_{\a, \alpha} (x)$ and $\overline{\Delta}^{\a, \beta}_\gamma (x)$ which describe the quark-antiquark systems bound by the strong interaction $V_{\alpha \gamma}^{\delta \beta}$ must be the solutions of the field equations (16). 

Among the most general bosonic field $\Delta_{\a, \alpha}^\gamma(x)$, let us consider  now a special class  of the scalar $\Delta_{\a}(x)$  by making the  projection 

$$\Delta^\gamma_{\a, \alpha} (x) = \left({1+\gamma_5\over 2}\right)^\gamma_{\alpha}\Delta_{\a}(x)\;,$$
\begin{equation}
\overline{\Delta}^{\a, \alpha}_\gamma (x) = \left({1-\gamma_5\over 2}\right)
_\gamma^{\alpha}\Delta^*_{\a}(x)\;.
\end{equation}

In some sense, as we will see, this composite scalar $\Delta_{\a}(x)$ 
substitutes the standard elementary Higgs field to generate masses to both 
the top quark  
and the gauge bosons.  
Associated to this particular $\Delta_{\a}(x)$ case, the functionals (12) and (13) become
\begin{equation}
W^{(2)} \left[ \Delta,\overline{\Delta}\right] = \I \int \d^4 x \; \d^4 y \,
\Delta^*_{\a} (x) \Delta_{\a} (y)
\lambda(y-x)\;,
\end{equation}
$$W^{(4)} \left[ \Delta,\overline{\Delta}\right] = {\I\over 2} \int \d^4 x\, \d^4 y \,\d^4 z \, \d^4 w \,\Delta^*_{\a} (x) \Delta_{\a} (y)
$$
\begin{equation}
\times\;\Delta^*_{\b} (z) 
\Delta_{\b} (w)\, \Pi(x-y, y-z, z-w) \;,
\end{equation}
where
\begin{equation}
\lambda(x-y) = {\rm Tr}\left[S^L(x-y)S^R(y-x)\right] \;,
\end{equation}
\begin{equation}
 \Pi(x-y, y-z, z-w) ={\rm Tr} \left[S^L(x-y)S^R(y-z)S^L(z-w)S^R(w-x)\right] \;.
\end{equation}
For the other functionals $W^{(2n)}\left[ \Delta,\overline{\Delta}\right]$, $n>2$, we have similar expressions easily generalized.  They are all nonlocal functionals of the scalar fields $\Delta_{\a} (x)$ and $\Delta^*_{\a} (x)$. Each of them can be expressed in terms of a corresponding local functional of $\Delta_{\a} (x)$ and $\Delta^*_{\a} (x)$ and their derivatives $\partial^\mu \partial^\nu \cdots \partial^\lambda
\Delta_{\a} (x)$, $\partial^\mu \partial^\nu \cdots \partial^\lambda
\Delta^*_{\a} (x)$ to all orders.

In the presence of the vector gauge boson fields 
$\left[W_\mu (x)\right]^{\b}_{\a}$ and $B_\mu (x)$, the ordinary derivatives 
$\partial_\mu$ must be replaced by the covariant ones $\D_\mu$, thus 
\begin{equation}
\partial_\mu \Delta_{\a} (x) \longrightarrow \D_\mu \Delta_{\a} (x) =
\partial_\mu \Delta_{\a} (x) +\I \left[A_\mu (x)\right]^{\b}_{\a} \Delta_{\b} (x)\;,
\end{equation}
\begin{equation}
\left[A_\mu (x)\right]^{\b}_{\a} = g \left[W_\mu (x)\right]^{\b}_{\a} + {g'\over 2} B_\mu(x) \delta^{\b}_{\a} \;.
\end{equation}
Up to the second order with respect to the first derivatives
$\D_\mu \Delta_{\a} (x)$, $\D_\mu \Delta^*_{\a} (x)$ and the first order with respect to the second derivatives $\D_\mu\D_\nu \Delta_{\a} (x)\;,$  $\,\D_\mu\D_\nu \Delta^*_{\a} (x)$, we obtain   

\begin{equation}
W^{(2)} \left[ \Delta,\overline{\Delta}\right] = \I \int \d^4 x \left\{ 
\lambda ^{(0)}(x)\Delta^*_{\a} (x) \Delta_{\a} (x) +{1\over 2} \lambda^{(2)}(x)\Delta^*_{\a} (x)  D^\mu  D_\mu \Delta_{\a} (x)\right\}\;,
\end{equation}
$$W^{(4)} \left[ \Delta,\overline{\Delta}\right] = {\I\over 2} \int \d^4 x \left\{ \Pi^{(0)}(x)\Delta^*_{\a} (x) \Delta_{\a} (x) \Delta^*_{\b} (x) \Delta_{\b} (x) \right.
$$
\begin{equation}
+ \left. \Pi ^{(2)}(x) \Delta^*_{\a} (x) 
\Delta_{\a} (x) \Delta^*_{\b} (x)  D^\mu  D_\mu  
\Delta_{\b} (x) +  \Omega (x)\Delta^*_{\a} (x)  D^\mu \Delta_{\a} (x)
 \Delta^*_{\b} (x)  D_\mu   
\Delta_{\b} (x) \right\}\;,
\end{equation}

where
\begin{equation}
\lambda ^{(0)}(x) =\int \d^4 y \,\lambda(x-y)\;,
\end{equation}
\begin{equation}
\lambda^{(2)}(x) =\int \d^4 y \,(y-x)^\mu \,(y-x)_\mu \lambda(x-y)\;,
\end{equation}
\begin{equation}
\Pi^{(0)}(x) =\int \d^4 y \, \d^4 z \,\d^4 w \, \Pi(x-y, y-z,z-w)\;,
\end{equation}
\begin{equation}
\Pi^{(2)}(x) =\int \d^4 y \,\d^4 z \,\d^4 w \,(y-x)^\mu (y-x)_\mu \Pi(x-y, y-z,z-w)\;,
\end{equation}
\begin{equation}
\Omega (x) =\int \d^4 y \,\d^4 z \,\d^4 w \,(y-x)^\mu (w-z)_\mu \Pi(x-y, y-z,z-w)\;.
\end{equation}

For $W^{(2n)} \left[ \Delta,\overline{\Delta}\right]$ with $n>2$, we have expressions generalizing (24) and (25). 

Now in both sides of the field equations (16), as well as in (24), (25) and their generalizations for $W^{(2n)} \left[ \Delta,\overline{\Delta}\right], n>2$, we set the scalar composite fields $\Delta_{\a} (x)$ to be equal to their vacuum expectation values $\Delta^{(0)}_{\a}$ which describe the condensate of quark-antiquark pairs. From (1) and (2), we note that among the two components $\Delta^{(0)}_{\a}$ of the doublet, only the second component  $\Delta^{(0)}_{\a=2}$  is a neutral field  and could have non-vanishing 
vacuum expectation value, thus
\begin{equation}
\Delta^{(0)}_{\a} =\delta_{\a 2} \Delta
\end{equation}
where $\Delta$ is some real constant that we put equal to the top mass
$\Delta = M_t$. This point can be seen from the term $\overline{\Delta}_\gamma^{\a, \beta}(x)
 \psi_{\a, \beta} (x) \overline{\chi}^\gamma (x) + \chi_\delta (x) 
\overline{\psi}^{\a, \alpha} (x) \Delta^\delta_{\a, \alpha} (x)$ figured on the second line of (8).
With this constant value (31) of the scalar fields $\Delta_{\a} (x)$, the field equation (16)  reduces to
\begin{equation}
{1\over 4 G} = {N_c^2-1\over 2N_c} {-\I \over (2\pi)^4} \int {\d^4 p \over p^2 -M^2_t +\I\epsilon }\;,
\end{equation}
where the implicit number of quark colors $N_c=3\,$ is now taken into account 
in  $V^{\delta \beta}_{\alpha \gamma}$ of (4) and (16).
The divergence of the integral in the r.h.s. of (32) may be handled by a momentum cutoff $\Lambda$ written in a covariant manner, and we obtain the algebraic equation
\begin{equation}
 {N_c^2-1\over 2N_c}{1\over 4 \pi^2}\int_0^{\Lambda^2} { x \,\d x\over x+ M^2_t} =
{N_c^2-1\over 2N_c}{1\over 4\pi^2 }
\left[\Lambda^2 -M^2_t \, \ln {\Lambda^2 +M^2_t \over M^2_t }\right]
 ={1\over G}\;.
\end{equation}
The solution of this equation does exist if and only if
$\Lambda$ satisfies the condition
\begin{equation}
 \Lambda^2 > {4\pi^2\over G } {2N_c\over N_c^2-1} \;.
\end{equation}
From (33), let us denote by $y=f(x)$ the positive solution of the equation $ \ln (1+y)=xy$ in the interval $0<x<1$, then we have
\begin{equation}
 M_t^2 ={\Lambda^2 \over f(x_0) }\;\;,\; x_0 = 
1 -{4\pi^2 \over G \Lambda^2}{2 N_c \over N_c^2-1}\;.
\end{equation}
This equation, reminiscent of \cite{njl}, determines $ M_t^2$ in terms of
$\Lambda^2 $ and $G$.

In order to derive the gauge boson  masses, we substitute the constant value (31) of the scalar fields $\Delta_{\a}(x) $ into the expressions of $W^{(2n)} \left[ \Delta,\overline{\Delta}\right]$, sum up all these expressions 
and separate out the quadratic term  $A_\mu A^\mu$ which contributes to the vector gauge boson mass  in  the effective Lagrangian ${\cal L}^{A}_{mass}(x)$. We finally obtain

\begin{equation}
{\cal L}^{A}_{mass}(x) ={-1\over 2} \left\{ M_t^2 \; {\cal I} \;
\left[A_\mu (x)\right]_2^{\c} \left[A^\mu (x)\right]^2_{\c}
 +  M_t^4 {\cal J} \left[A_\mu (x)\right]_2^{2} \left[A^\mu (x)\right]^2_{2}
\right\}\;,
\end{equation}
where
\begin{equation}
{\cal I} = -{\I\over (2\pi)^4} \int \d^4 p { Tr \left[\widetilde{S}^L(p) 
\widetilde{S}^R(p)\widetilde{S}^L(p) 
\widetilde{S}^R(p) \right] \over 1- { M_t^2 \over p^2}}\;,
\end{equation}

\begin{equation}
{\cal J} = {-1\over 2} {\I \over (2\pi)^4} \int \d^4 p 
{ Tr \left [\widetilde{S}^L(p) 
\widetilde{S}^R(p) \widetilde{S}^L(p) 
\widetilde{S}^R(p) \widetilde{S}^L(p) 
\widetilde{S}^R(p) \right]\over \left( 1- { M_t^2 \over p^2}\right)^2}\;.
\end{equation}

 $\widetilde{S}^L(p)$ and 
$\widetilde{S}^R(p)$ being the propagators of the free left-handed and right-handed massless fermions in momentum space:
\begin{equation}
\widetilde{S}^L(p) = {1-\gamma_5\over 2} \;{\I \psl\over p^2 +\I\epsilon }\;\;,\;\;
\widetilde{S}^R(p) = {1+\gamma_5\over 2} \;{\I \psl\over p^2 +\I\epsilon }\;.
\end{equation}
Since  ${\cal I}$ is divergent, we again use the cutoff $\Lambda$ in a covariant manner and get  
\begin{equation}
{\cal I}= {N_c \over 8 \pi^2}\int_0^{\Lambda^2} { \d x\over x+ M^2_t} =
{N_c\over 8\pi^2 }
\ln {\Lambda^2 +M^2_t \over M^2_t } \;.
\end{equation}
The integral ${\cal J}$ is convergent and equals
\begin{equation}
{\cal J}= -{N_c \over 16 \pi^2}\int_0^{\infty} { \d x\over (x+ M^2_t)^2} =
-{N_c \over 16\pi^2 }
{1\over M^2_t } \;.
\end{equation}
Setting
\begin{equation}
\left(W_\mu\right)^1_2 ={1\over \sqrt{2}} W_\mu^+ \;\;,\;\;\left(W_\mu\right)_1^2 ={1\over \sqrt{2}} W_\mu^- \;,
\end{equation}
\begin{equation}
\left(W_\mu\right)^1_1 = - \left(W_\mu\right)_2^2 ={1\over 2}\left[ \sin \theta_W A_\mu +\cos \theta_W  Z_\mu\right] \;,
\end{equation}
\begin{equation}
B_\mu = \cos \theta_W A_\mu -\sin \theta_W  Z_\mu \;,
\end{equation}

and using $ g\sin \theta_W - g'\cos \theta_W  =0$, we obtain from (36)
\begin{equation}
{\cal L}^{A}_{mass}(x)  =-M_W^2 \,W_\mu^+(x)W^\mu (x) -{1\over 2} M_Z^2 \,Z_\mu(x)Z^\mu (x)\;,
\end{equation}
where the gauge boson masses are found to be
\begin{equation}
M_W^2 = N_c {g^2\over 32 \pi^2} M_t^2 \ln {\Lambda^2 +M^2_t \over M^2_t }\;,
\end{equation}
\begin{equation}
M_Z^2 = N_c {g^2 +g^{'2}\over 32 \pi^2} M_t^2 \left[\ln {\Lambda^2 +M^2_t 
\over M^2_t } -{1\over 2}\right] \;.
\end{equation}
Equation (46) is reminiscent of the Pagel-Stokar formula\cite{ps} which relates the charged pion decay constant $f_\pi \approx \; 131 \;{\rm MeV}$ 
to the dynamically generated  mass obtained in QCD for the up down quarks. From QCD to electroweak interactions, the  $f_\pi$ of the former is replaced by the vacuum expectation value $v = [\sqrt{2} G_{\rm F}]^{-1/2}  = 2M_W/g \approx 246$ GeV of the latter. This may be schematically transcribed as: 

In QCD, the Pagel-Stokar formula relates $f_\pi $ to $m_{\rm u,d}$. In dynamical symmetry breaking of the electroweak interactions, equation (46) gives $v$ in terms of $ M_t$.
 
Furthermore, with $g^2+g^{'2} = g^2/\cos^2 \theta_W$, we get from (46) and (47)
\begin{equation}
 \rho \equiv {M_W^2 \over M_Z^2 \cos^2 \theta_W} = 1+ {1\over 2 \ln 
\left({\Lambda^2  \over M^2_t }+1\right)} \;.
\end{equation}
The precision electroweak measurements at the $10^{-3}$ level force the scale $\Lambda$ of top-condensation models to be very high about $10^{15}$ GeV when we  compare experimental data with (46), (47) and (48).
 
It is gratifying to note that using very different 
methods, we recover some results previously obtained in the literature
\cite{Bar, njl, ps}. 
We also remark that in the standard Higgs mechanism, at the tree-level the parameter $\rho$ is equal to unity. The correction $\Delta \,\rho =|\rho-1|$ can only come from higher order loops to which the top quark contribution at order $g^2$ is quadratic  in  its mass  $M_t$, thus\cite{hkp} 
\begin{equation}\Delta \rho = {3 g^2  \over 64\pi^2 \cos^2\theta_W  }\, {M_t^2 \over M_W^2}\;.
\end{equation}
The result (49) is in sharp contrast with the dynamical symmetry model discussed
 here for which the $\rho$ parameter is already different from unity to  zero order of the coupling constants $g, g'$, as shown by equation (48). The unbroken global symmetry of the Higgs sector (translated by $\rho=1 $) usually called custodial $SU(2)$ symmetry\cite{SSVZ} of the  electroweak theory  is systematically 
violated  by a smooth logarithm of $M_t$. The patterns of  custodial $SU(2)$ symmetry breaking as illustrated by (48) and (49) are conceptually not the same. 
 
We now briefly discuss some phenomenological consequences of the top-condensation models, in particular the production and decay modes of the neutral scalar denoted by $H_t$ which replaces the elementary Higgs boson $H^0$ of the standard model. 

1- In general, whatever the schema invoked to dynamically generate the top and the $W, Z$ masses, there must exist a triplet of new Nambu-Goldstone bosons  resulting from the breaking of chiral symmetry in the top-bottom system postulated in (1) $\cdots$ (4). They are absorbed by the $W,Z$ to acquire masses as shown in (46), (47). Furthermore, a neutral CP-even state analogous to the $\sigma$ boson in  QCD must exist, it is a scalar $\overline{t} t$ bound state denoted by  $H_t$. Since the 
force postulated in (4) that ties top-antitop pair is so strong that it can dynamically generate such a huge $175$ GeV mass, it is likely that the biding energy is large in $H_t$ and its mass could be smaller\cite{ Del} than twice\cite{Nambu} the top mass. This is taken as a very rough indication of where to find  $H_t$.
 
2- The production cross section of $H_t$ is governed by gluon-gluon fusion\cite{hkp}
 into $t\overline {t}$ pairs, so that hadron colliders at the FermiLab Tevatron and the Cern LHC are appropriately the right places for $H_t$ searches. On the other hand, the  $ Z+ H_t$ associated production  by lepton colliders,  $e^+ e^- \to Z^* \to Z + H_t  $ is largely suppressed, since direct coupling $ZZ H_t$ is absent. This is also in sharp contrast with the elementary Higgs boson $H^0$ production dominated\cite{hkp} by the reaction $e^+ e^- \to Z^* \to Z + H^0 $, because direct coupling $ZZ H^0$ is large.

3- Due to the nature of a strongly  bound $\overline{t}t$ state, the $H_t$ is leptophobic, and "almost" hadrophobic, i.e. it cannot decay into leptons and "light" hadrons made up by the first two families up, down, charm, strange quarks at the tree level (lowest order of the coupling constants $g, g'$).  Indeed, the Yukawa direct couplings of $H_t$ with leptons and the first two families of quarks are absent, contrarily to the standard elementary Higgs boson $H^0$ case. Therefore the $H_t$ although so massive would have  a very narrow width (only a few GeV) in sharp contrast with  the standard elementary Higgs boson $H^0$ which has  a width\cite{hkp} around  18 GeV for  a  hypothetical 350 GeV $\approx 2M_t $ mass. The decays of $H_t$ into "light" hadrons can  only proceed  through gluons emission, similarly to quarkonium $J/\psi$ and $\Upsilon$ decays which are  suppressed by the Okubo-Zweig-Iizuka (OZI) rule reflecting QCD asymptotic freedom. 

4- The most distinctive signatures of $H_t$ would be its dominant decay modes into the bottomonium $\Upsilon$ and an energetic photon or $\Upsilon$ accompanied by the $Z$ weak boson, as depicted by Fig.1. This comes from the special situation of the third family top-bottom quarks which have the additional topcolor interaction -- mediated
 by topgluons ${\cal G}_t$ -- thus destroying the universality between the three quark families. Due to  the non-universal character of the top-bottom system which does not possess the Glashow-Iliopoulos-Maiani (GIM) cancelation, flavor changing neutral decay of
 $H_t$ into $t \overline{c} +\overline{t}c$ channels is another
 spectacular signature\cite{bur} of the top condensation models.  

The ratio $ \Gamma(H_t \to \Upsilon +Z)/\Gamma(H_t \to \Upsilon +\gamma
)$ is found to be
\begin{equation}
{\Gamma(H_t \to \Upsilon +Z)\over \Gamma(H_t \to \Upsilon +\gamma
)} =\left ({3\over 4\sin\theta_W \cos\theta_W} \right)^2\left[ \left({1\over 2}- {4\over 3}\sin^2\theta_W \right)^2 +{1\over 4} \right] \left( 1- {M_Z^2\over M^2_{H_t}} \right) \approx 0.8\;.
\end{equation}
 Finally we remark that hadronic decays of $H_t$  must proceed into b$\overline{\rm b}$ pair through two topgluons ${\cal G}_t$ exchange in Fig.2. These  dominant  decay  modes into $B^*,B $ mesons 
   $H_t \to B^*(B) +\overline{B}^*(\overline{B})$  are OZI unsuppressed.

\begin{em}

\end{em}


\begin{thebibliography}{50}
%
\bibitem{Higgs}
 P. W. Higgs, Phys. Lett. {\bf 12}, 132 (1964); Phys. Rev. Lett. {\bf 13}, 508
 (1964).      

\bibitem{Eng}
F. Englert and R. Brout,  Phys. Rev. Lett. {\bf 13}, 321
 (1964).

\bibitem{Gur}
G. S. Guralnik, C.R. Hagen and T.W.B. Kibble,  Phys. Rev. Lett. {\bf 13}, 585
 (1964).

\bibitem{Gsw}
The genesis and developments of the Electroweak Standard Model are nicely reviewed  in {\bf The rise of the Standard Model}, ed. L. Hoddeson, L. Brown, M. Riordan and M. Dresden, Cambridge University Press (1997). See in particular the lectures of S. Bludman, J. Iliopoulos, M. Kobayashi, D. Perkins, C. Prescott, S. Schweber, G.'t Hooft, M. Veltman and S. Weinberg.
       
\bibitem{Nambu}
       Y. Nambu in {\bf New Theories in Physics}, 
Proc. XI Warsaw Symposium on Elementry Particle Physics, pp. 1-10, ed. Z. Adjuk, S. Pokorski and A. Trautman,  World Scientific, Singapore (1989).

\bibitem{Mir}
V.A. Miransky, M. Tanabashi and M. Yamawaki, Phys.
Lett. {\bf B221}, 177 (1989).

\bibitem{Bar}
W.A. Bardeen, C. T. Hill and M. Lindner, Phys. Rev. {\bf D41}, 1647 (1990). 

\bibitem{yem}
Xuan Yem Pham, Phys. Lett. {\bf B241}, 111 (1990).

\bibitem{Hil}
C. T. Hill,  Phys. Lett. {\bf B345}, 483 (1995). 

\bibitem{Del}
D. Delepine, J.M. G\'erard and R. Gonzalez Felipe, Phys.
Lett. {\bf B372}, 271 (1996).

\bibitem{Chi}
R.S. Chivukula, B. Dobrescu, H. Georgi and C.T. Hill, Phys. Rev. {\bf D59}, 075003 (1999).

\bibitem{Cvetic}
       G. Cvetic, Rev. Mod. Phys. {\bf 71}, 513 (1999).


\bibitem{Hub}
J. Hubbard, Phys. Rev. Lett. {\bf 3}, 77
 (1954); R.L. Stratonovich, Sov. Phys. Solid State {\bf 2}, 1824
 (1958).

\bibitem{hieu}
        Nguyen Van Hieu in {\bf Computational Approaches for Novel Condensed Matter Systems}, pp. 194-234, ed. M. Das, Plenum Press, New York (1994);  Aus. J. Phys. {\bf 50}, 1035 (1997).

\bibitem{njl}
       Y. Nambu and G. Jona-Lasinio, Phys. Rev. {\bf 122}, 345 (1961).

\bibitem{ps}
H. Pagels and S. Stokar,  Phys. Rev. {\bf D20}, 2947 (1979).

\bibitem{hkp} See Standard Model textbooks, for instance {\bf Elementary Particles and Their Interactions, Concepts and Phenomena} pp.602-607, by Quang Ho-Kim and Pham Xuan Yem, Springer, Berlin  (1998).

\bibitem{SSVZ} P. Sikivie, L. Susskind, M. Voloshin and V. Zakharov, Nucl. Phys. {\bf B173}, 189 (1980).

\bibitem{bur}
      G. Burdman, Phys. Rev. Lett. {\bf 83}, 2888
 (1999).


%
\end{thebibliography}
\end{document}